\documentclass[aps,prd,showpacs,twocolumn,amssymb,floatfix,nofootinbib,superscriptaddress]{revtex4-1}
\usepackage{graphicx,amsmath,subfigure,psfrag,times}

\usepackage{amssymb}
\usepackage{graphicx}
\usepackage{epsfig}
\usepackage{amsfonts}
\usepackage{amssymb}
\usepackage{amsmath}
\usepackage{latexsym}
\usepackage{bm}
\usepackage{ulem}
\newtheorem{thm}{Theorem}

\newtheorem{df}[thm]{Definition}

\begin{document}
\title{Quantum phase transition from the topological viewpoint}

\author{Izumi Tanaka}
\email{tanaka-i@eng.setsunan.ac.jp}
\affiliation{%
 Academic Support Center, Setsunan University, Neyagawa 572-8508, Japan }%
 \date{\today}

\begin{abstract}        
This study targets 
quantum phases which are characterized by topological properties 
and not associated with the symmetry breaking. 
We concern ourselves primarily with
the transitions among these quantum phases. 
This type of quantum phase transition was investigated by $G$-cobordism in unified framework.  
This framework provides a useful method to investigate a new quantum phase.  

\end{abstract}


\maketitle

\section{Introduction}
In recent years, various fields have utilized ideas of topology 
where correlation among local and global nature can be brought out.
This trend has been applied to the areas of condensed matter as well \cite{Volovik2003}.
To understand the various phase of matter, concept of order parameter was essential in 
Landau-Ginzburg theory. 
Quantum phases in the absence of order parameter have drawn increasing attention 
after the discovery of quantum Hall phase and Haldane phase \cite{Klitzing-Dorda-Pepper1980, Haldane1983}.  

Analysis of the transitions among the quantum phases has revealed the several aspects \cite{Lohneysen-Rosch-Vojta-Wolfle2007}. 
First, certain class of quantum phase transition (QPT) has been understood by the mechanism of spontaneous symmetry breaking. A basic example is the order-disorder transition in the one-dimensional quantum Ising chain in a transverse field. 
Second, continuous QPT as Lifshitz transition can be adduced.  
This transition occurs in noninteracting Fermion systems, which is characterized by the topology change of the Fermion surface in the Brillouin zone \cite{Yamaji-Misawa-Imada2006}. 

QPT among quantum phases with topological orders 
can be point out as the third aspect. 
For example, consider the degeneracy of ground state depends on the topological 
configuration where the physical system lives.
Since the ground-state degeneracies of the states are robust against arbitrary perturbations,  
the phases with different ground-state degeneracies have different topological orders. 
Because of the different ground-state degeneracy, 
the transition from one ground state to another ground state may be not continuous, further it  
is not associated with a change in the symmetries of the states. It is suggested that measuring topological degeneracy is one of the simplest ways to probe the topological order in a system.
As the Hamiltonian is changed in the theory, the ground-state degeneracy may jump which signals a 
phase transition between two phases with different topological orders. 
 \cite{Wen1989, Wen-Niu1990}. 
%
We term this type of QPT topological quantum phase transition (TQPT).  

Transition connects quantum phases characterized by the topological invariants 
can be also regard as TQPT. 
For example, in the quantum Hall phase,  hall conductance of filled magnetic Bloch band is 
always defined in terms of the topological invariant, Chern number \cite{Klitzing-Dorda-Pepper1980,  Thouless-Kohmoto-Nightingale-Nijs1982, Lauglin1983, Simon1983, Avron-Seiler-Simon1983, Niu-Thouless-Wu1985}.
These phases are not associated with the symmetry breaking and 
do not have the local order. They are robust against perturbations and have topological order. 
%
When TQPT occurs, topological invariant is preserved totally and/or topology of internal space or ground state changes without symmetry breaking.


Several earlier studies have tried to explain these TQPT. 
Some studies on quantum Hall effect (QHE) have 
demonstrated that the band crossing effect based on the Dirac fermion model 
explain the TQPT 
\cite{Oshikawa1994, Hatsugai-Kohmoto-Wu1996, Bellissard1995, Lee-Chang1998}. 

Consequently, there have been a growing interest in understanding the phase of matter from the point of view of topology. To analyze the TQPT is imperative in the pursuit of understanding of the quantum phases.

%
%
In this letter, we are trying to capture an elementary mechanism of 
the TQPT. 
We consider the unified treatment 
of the physical systems with topological properties. 
 %
This paper is organized as follows. 
In \S 2, basic framework is specified.
In \S 3, specific examples of TQPT are presented. 
Finally, \S 4 is devoted to conclusions.

\section{Basic framework}\label{sec:}
In order to begin addressing the questions, we would like to identify the what we meant by the TQPT.   
Quantum phases are characterized by 
topological order. 
Mathematically, 
further to manifold describing 
internal space or ground state, 
Lie group providing topological invariant is necessary for characterizing them. 
When taken together, these phases are described by $G$-manifold (compact low-dimensional manifold with a group action ). 
Not only turning our attention to changes of topological invariant, we also take into account the fact that 
change of topologies of space without the symmetry breaking.   
As a consequence, we investigate 
processes where topologies of the manifold and/or topological invariants change without symmetry breaking.   
$G$-cobordism is better suited for describing TQPT.

First, the characteristics of $G$-manifold are explained:  
 $G$-manifold 
has a fiber bundle structure, which is characterized as follows.
\begin{thm}
Let $G$ be a compact Lie group and $\widetilde{M}$ be $G$-manifold.

If $G$ action is free, orbital space $\widetilde{M}/G$ will serve as a manifold, and  $\pi: \widetilde{M}\mapsto \widetilde{M}/G$ becomes projection map of 
$C^{\infty}$ principal fiber bundle where $G$ is the fiber.

The differentiation structure on $\widetilde{M}/G$ is unique and fulfills the following conditions:

(i) $\pi: \widetilde{M}\mapsto \widetilde{M}/G$ is $C^{\infty}$ map.

(i\hspace{-.1em}i) $h:\widetilde{M}/G \mapsto \widetilde{M}$ is $C^{\infty}$ map if and only if $h \circ \pi$ is $C^{\infty}$ map.

\end{thm}
Let $G$ and $G_1$ be  compact Lie groups such that 
they admit orientations
that are preserved  by both left and right translations,
and that the skeletons of their classifying spaces are finite. 
Let $\widetilde{M}$ and $\widetilde{M}_1$   be an oriented compact differentiable
$n$-dimensional $G$-manifold without boundary.

Suppose that the topology of a $G$-manifold $\widetilde{M}$ changes to
 a $G$-manifold $\widetilde{M}_1$
\begin{equation}
(\widetilde{M},G) \longrightarrow (\widetilde{M}_{1}, G). 
\end{equation}
Note that the quotient map
$\widetilde{M} \mapsto \widetilde{M}/G$ is a principal $G$-bundle.
This topology change holds if and only if the bordism Stiefel-Whitney number $w_{I,x}(\widetilde{M},G)$ and bordism Pontrjagin number $p_{I,y}(\widetilde{M},G)$ are equal before and after the topology change.

We use the notation $\left[  \widetilde{M}, G \right]$ which denotes equivalence class ($G$-cobordant class) 
of $G$-manifold $ \widetilde{M}$. 
The equivalent relation is established by equal bordism Stiefel-Whitney number and bordism Pontrjagin number. Using this notation, above relation can be expressed as 
\begin{equation}
 (\widetilde{M},G), \hspace{1mm} (\widetilde{M}_{1}, G) \in  \left[  \widetilde{M}, G \right]   
\end{equation}
or 
\begin{equation}
 \left[  \widetilde{M}, G \right]   \sim  \left[  \widetilde{M}_1, G \right].  
\end{equation}

Above expression shows a cobordantness between $M=\widetilde{M}/G$ and  $M_1(=\widetilde{M}_1 /G_1)$:
$\partial {}^{\exists}W =M\sqcup M_1$.  
All spaces are oriented, since $G$ preserves the orientation. 
Basic points are given in appendix and the author's previous paper \cite{TN2011}. 
\section{Topological quantum phase transitions}\label{sec:}
TQPT without topology change can be expressed as follows by the $G$-cobordism: 
\begin{equation}
 \left[  \widetilde{M}_1\sharp \widetilde{M}_2\sharp \cdots \sharp \widetilde{M}_m , G \right]   \sim  \left[  \widetilde{M}_1\sharp \widetilde{M}_2\sharp \cdots \sharp \widetilde{M}_n, G \right]   \nonumber
\end{equation}
is equivalent to equality of topological invariant. 
Equality of bordism Pontrjagin and bordism Stiefel-Whitney number in lower dimension suggest the following. 
\begin{equation}\label{2d}
\int_{ M_1\sharp M_2 \sharp \cdots \sharp M_m } c=\int_{ M_1\sharp M_2 \sharp \cdots \sharp M_n }c  
\end{equation}
where $c$ is the Chern class determined by the group $G$, further $M_1=\widetilde{M}_1/G$ and $M_2= \widetilde{M}_2/G$. 
This includes the sum rule for the TKNN integers when two bands collide.
In fact, this is shown by $ \int_{ M_1\sharp M_2 } c=\int_{ M_1\sharp M_2 }c  $, 
where 
$M_{1}$ and $M_{2}$ represent adjacent energy bands, and $G=\mathbf{U}(1)$ 
derives from the Berry phase 
 \cite{Simon1983, Avron-Seiler-Simon1983, Oshikawa1994, Hatsugai-Kohmoto-Wu1996}.

Next we consider different situation. 
It is shown that the ground-state degeneracy depends on
the space topology. For example, on a Riemann surface with genus $g$ the 
vacuum degeneracy of chiral spin state is $2k^g$. $k$ is determined by $B=2\pi k$ where $B$ 
is uniform magnetic field \cite{Wen1989, Wen-Niu1990}. 
Accordingly, topology change of the space suggests the transition 
among states with different topological degeneracies. 
The primary consideration should be 
the effect of topology change of the space without symmetry breaking. 
We consider mainly
the system with lower dimension and gauge group such as $\mathbf{U}(1)$,  $\mathbf{SU}(2)$ or $\mathbf{SU}(3)$.

{ \it Change of topology in two or four dimension}  \hspace{2mm}
We consider the two-dimensional or four-dimensional manifolds such as $\widetilde{M}/G$ equals to $S^2$, $S^4$or $T^2$.
We explain necessary and sufficient condition the for following  
$\left[  \widetilde{M}, G \right]   \sim  \left[  \widetilde{M}_1, G \right]  $ where $M=\widetilde{M}/G $ and $M_1=\widetilde{M}_1/G $.

Equality of  both bordism Pontrjagin and bordism Stiefel-Whitney number is equivalent to the following  
\begin{align}
           \int_M c_1 = \int_{M_1}c_1, & & \hspace{1mm} (\text{two dimension, } G=\mathbf{U}(1))  \nonumber    \\
           \int_M c_2 =\int_{M_1}c_2,  & & \hspace{1mm}  (\text{four dimension, } G=\mathbf{SU}(2), \mathbf{SU}(3))  \nonumber    
\end{align}
where $c_1$ and $c_2$ is the first Chern class and second Chern class respectively. 
This suggests that topology change requires the constant topological invariant.

{\it Change of topology in three dimension}  \hspace{2mm}
Both of equality of bordism Pontrjagin number and bordism Stiefel-Whitney number
do not impose a limitation on the cobordism 
for $G=\mathbf{U}(1)$, $\mathbf{SU}(2)$, or $\mathbf{SU}(3)$ because of their dimensions.

{ \it Change of topology and topological invariant}   \hspace{2mm}
We investigate the transitions among different topological degeneracies and topological invariants. 
We base on the following process. 
\begin{equation}\label{Null}
\left[  \widetilde{M}, G_1 \times G_2\right]   \sim  \left[  \widetilde{M}', G_1 \right]  \times \left[ G_2, G_2 \right].  
\end{equation}
This holds if all bordism Pontrjagin number and bordism Stiefel-Whitney of
 $\left[ \widetilde{M}, G_1 \times G_2 \right]$ 
equals to zero, 
because $G_2/G_2$ is a point \cite{TN2011}.
This transition suggest the topology change of space and change of gauge group.
After the transition, there is no limit in topological invariant in the space $ M'= \widetilde{M}'/G_1$.

It is known that the ground-state degeneracy is unchanged after connecting $g$ tori by 
tubes to form genus $g$ Riemann surface for fractional quantum Hall (FQH) system. 
Each torus will generate a Heisenberg algebra. The algebra generated by different torus commute 
with each other. As the ground-state degeneracy on the torus is $q$-fold,  
total ground-state degeneracy of genus 
$g$ Riemann surface is $q^g$ \cite{Wen-Niu1990}.
In this case, count of degeneracy suggests that the genus 
$g$ Riemann surface just corresponds to the $T^2 \times \cdots \times T^2$ ($g$-folds), where $T^2$ is two-torus.
From this viewpoint, we consider following type of QPT. 
Applying Eq.(\ref{Null}), following type QPT is possible:
\begin{align}
\left[  \widetilde{M_1}\times \cdots \times \widetilde{M_n}, G_1 \times \cdots \times G_n\right]  \sim & & \nonumber \\
\left[  \widetilde{M}_1\times \cdots \times \widetilde{M_i}, G_1 \times \cdots \times G_i \right]  &\times & \left[  \widetilde{M}_{i+1}, G_{i+1} \right] \hspace{8mm} \nonumber \\
&\times & 
 \cdots \times  \left[  \widetilde{M}_n, G_n \right],  \nonumber
\end{align}
where $1 < i \le n. $
This holds if all bordism Pontrjagin number and bordism Stiefel-Whitney of $\left[ \widetilde{M}_1\times \cdots \times \widetilde{M}_n, G_1 \times \cdots \times G_n \right]$ equals to zero 
if some $\widetilde{M}_i/G_i$ is a point. 
Also in this case, 
after the transition, arbitrary topological invariant is possible in the space $ \widetilde{M}_1\times \cdots \times \widetilde{M_i}/G_1 \times \cdots \times G_i$. 
In above process, $\widetilde{M_k}/G_k=T^2$, $(1\le k\le i)$ may correspond to 
genus $i$ Riemann surface.

\section{Conclusions}\label{sec:Sum}

This letter finds the significance of TQPT by unified treatment. 
Through TQPT, change of topology and/or topological invariant is considered.  
In this process, gauge group can change without symmetry breaking.  
Various quantum phases can be established from TQPT.

\appendix
\section{ G-cobordism}

\begin{df}
 Let $g:N \mapsto Y$ be a singular $n$-manifold of
 a topological space $Y$.
For  $g:N\mapsto Y$,
 a partition of some natural number $I=(k_{1}, k_{2}, \cdots, k_{m})$,
and cohomology classes $x\in H^{*}(Y; \mathbb{Z}_{2}), y\in H^{*} (Y; \mathbb{Z})$,

set
\begin{align}
w_{I, x}(g) & =\langle w_{k_{1}}(N) \cdots w_{k_{m}}(N) (g^{*}x)
\hspace{0.5mm}, [N]_{2} \rangle\in\mathbb{Z}_{2}\nonumber\\
p_{I, y}(g)  & =\langle p_{k_{1}}(N) \cdots p_{k_{m}}(N) (g^{*}y)
\hspace{0.5mm}, [N] \rangle\in\mathbb{Z}.\nonumber
\end{align}
These are called bordism Stiefel-Whitney number and bordism Pontrjagin number of
$g$  for $I$ respectively .

If $f_{G}$ is the classifying map  
$f_{G}: \widetilde{M}/G\mapsto \mathbb{B}G$
 of a free $G$-action $\psi : \widetilde{M}\times G\mapsto \widetilde{M}$, then
 $w_{I,x}(f_{G})$ and $p_{I,y}(f_{G})$ are denoted by
 $w_{I,x}(\widetilde{M}, G)$ and $p_{I,y}(\widetilde{M}, G)$, and are called
 the bordism Stiefel-Whitney number and bordism Pontrjagin number 
 of $\psi $
 respectively.
\end{df}

\begin{thm} \label{G-cobordism}
Let $\widetilde{M}$  be an $n$-dimensional oriented compact
$G$-manifold without boundary.
Suppose that the homology group $H_{*} (\mathbb{B}G)$ has no torsion and that the Thom
homomorphism $\mu: \Omega_*(\mathbb{B}G)\mapsto H_{*}(\mathbb{B}G)$ is  surjective.
 Then
$\left[ \widetilde{M}, G\right] =0\in\Omega_{n} (G:\mathfrak{F}%
_{1}) $ 
 if and only if all of the bordism Pontrjagin numbers and the bordism
Stiefel-Whitney number of the $G$-manifold vanish.
\end{thm}


\bibliography{<your-bib-database>}



\end{document}